\documentclass[pdflatex,sn-nature]{sn-jnl}
\usepackage{graphicx}%
\usepackage{multirow}%
\usepackage{amsmath,amssymb,amsfonts}%
\usepackage{amsthm}%
\usepackage{mathrsfs}%
\usepackage[title]{appendix}%
\usepackage{xcolor}%
\usepackage{textcomp}%
\usepackage{manyfoot}%
\usepackage{booktabs}%
\usepackage{algorithm}%
\usepackage{algorithmicx}%
\usepackage{algpseudocode}%
\usepackage{listings}%
\usepackage{soul} 
\usepackage{xr} 
\usepackage{geometry}
\theoremstyle{thmstyleone}

\theoremstyle{thmstyletwo}%
\theoremstyle{thmstylethree}%
\newcommand{\ELvar}{\lambda_{\mathrm{Var}}}
\newcommand{\ELac}{\lambda_{\mathrm{AC1}}}
\newcommand{\EAC}{\widehat{\mathrm{AC1}}}
\newcommand{\EVar}{\mathrm{Var}[X]}
\newcommand{\dint}{\mathrm{d}}
\begin{document}

\title[Article Title]{The influence of data gaps and outliers on resilience indicators}

\author*[1,2,3]{\fnm{Teng} \sur{Liu}}\email{teng.liu@tum.de}
\author*[2,3]{\fnm{Andreas} \sur{Morr}}\email{andreas.morr@tum.de}
\author[2,3]{\fnm{Sebastian} \sur{Bathiany}}
\author[2,3]{\fnm{Lana L.} \sur{Blaschke}}
\author[2,3]{\fnm{Zhen} \sur{Qian}}
\author[2,4]{\fnm{Chan} \sur{Diao}}
\author[5]{\fnm{Taylor} \sur{Smith}}
\author[2,3]{\fnm{Niklas} \sur{Boers}}

\affil[1]{\orgdiv{School of Systems Science and Institute of Nonequilibrium Systems}, \orgname{Beijing Normal University}, \city{Beijing}, \postcode{100875}, \country{China}}
\affil[2]{\orgdiv{Earth System Modelling}, \orgname{School of Engineering and Design, Technical University of Munich}, \city{Munich}, \postcode{80333}, \country{Germany}}
\affil[3]{\orgname{Potsdam Institute for Climate Impact Research}, \city{Potsdam}, \postcode{14473}, \country{Germany}}
\affil[4]{\orgname{Faculty of Geographical Science},\orgname{Beijing Normal University}, \city{Beijing}, \postcode{100875}, \country{China}}
\affil[5]{\orgname{Institute of Geosciences},\orgname{Universit{\"a}t Potsdam}, \city{Potsdam}, \postcode{14476}, \country{Germany}}

\abstract{
The resilience, or stability, of major Earth system components is increasingly threatened by anthropogenic pressures, demanding reliable early warning signals for abrupt and irreversible regime shifts. Widely used data-driven resilience indicators based on variance and autocorrelation detect `critical slowing down', a signature of decreasing stability. However, the interpretation of these indicators is hampered by poorly understood interdependencies and their susceptibility to common data issues such as missing values and outliers. Here, we establish a rigorous mathematical analysis of the statistical dependency between variance- and autocorrelation-based resilience indicators, revealing that their agreement is fundamentally driven by the time series' initial data point. Using synthetic and empirical data, we demonstrate that missing values substantially weaken indicator agreement, while outliers introduce systematic biases that lead to overestimation of resilience based on temporal autocorrelation. Our results provide a necessary and rigorous foundation for preprocessing strategies and accuracy assessments across the growing number of disciplines that use real-world data to infer changes in system resilience.
}
\maketitle

\section{Introduction}
Resilience is a fundamental property of dynamical systems with stable equilibrium states, describing their ability to absorb disturbances and recover from perturbations without undergoing fundamental shifts in structure or function~\cite{peterson1998ecological,bathiany2024ecosystem}. In systems with strong nonlinear interactions and resulting positive feedbacks, small perturbations may trigger irreversible transitions if resilience is low~\cite{scheffer2001catastrophic}. This behaviour is observed across diverse domains, from climate systems and financial markets to biological networks and ecosystems~\cite{boers2022theoretical,battiston2016complexity,gao2016universal,folke2004regime}.

Monitoring the resilience of nonlinear Earth system components is crucial, particularly in the face of intensifying anthropogenic pressures. Several major climate subsystems have been suggested to be at risk of critical transitions in response to anthropogenic forcing \cite{boers2022theoretical,lenton2024remotely}, but anthropogenic forcing can also trigger abrupt responses in other parts of the Earth system, such as ecosystems. Ecosystem resilience plays a fundamental role in maintaining biodiversity, natural carbon sinks, and other essential ecosystem services~\cite{rockstrom2021we}. However, anthropogenic climate change and human activities are increasingly eroding this resilience, potentially pushing many ecosystems toward critical thresholds~\cite{moore2022getting,forzieri2022emerging}. A prominent example is the Amazon rainforest, where positive atmosphere-vegetation feedbacks and repeated disturbances such as droughts, wildfires, and deforestation are weakening the ability of forests to maintain their current state, raising concerns about an abrupt shift to a drier, less biodiverse, and low-tree-cover system~\cite{boers2017deforestation,bathiany2024resilience,blaschke2024spatial,zemp2017deforestation,lovejoy2018amazon,bochow2023south,flores2024critical}. Similar decreases in resilience also threaten coral reefs, boreal forests, and tundra systems, where external shocks can lead to long-term transformation in ecosystems~\cite{lenton2024remotely}.

Understanding and quantifying resilience is essential for predicting and mitigating abrupt regional shifts. Empirical approaches typically define resilience as the rate of recovery from external disturbances~\cite{pimm1984complexity,lenton2022resilience}. Consequently, measuring resilience directly requires either controlled experiments involving artificial perturbations or natural observations of strong external disturbances~\cite{dakos2022ecological}. An alternative approach is based on the fluctuation-dissipation theorem, which states that the rate at which a system returns to equilibrium after a disturbance can be inferred from its internal variability~\cite{kubo1966fluctuation}. This allows one to leverage statistical indicators to infer resilience from small, natural fluctuations within a system~\cite{smith2022empirical}. Key statistical measures derived from this principle include variance (representing the magnitude of fluctuations) and lag-one autocorrelation (AC1, a measure of the system's memory) ~\cite{carpenter2006rising,held2004detection}. As a system approaches a critical threshold, its recovery rate decreases, a phenomenon referred to as critical slowing down (CSD). Consequently, both variance and AC1 tend to increase~\cite{scheffer2009early,dakos2012robustness}. These CSD indicators have been widely used as proxies for resilience changes and as early warning signals of abrupt change in various types of natural systems, including ecosystems~\cite{boulton2022pronounced,smith2022empirical,forzieri2022emerging}, climate systems~\cite{boers2021observation,liu2023teleconnections,ben2023uncertainties}, and palaeoclimate records~\cite{boers2018early,boers2021critical,dakos2008slowing}. 

Despite broad applicability, CSD-based resilience indicators face several challenges when they are inferred from real-world data and sometimes produce false alarms~\cite{wagner2015false,smith2023reliability_ESD,ben2023uncertainties,ben2024uncertainties}. For instance, multi-sensor data with varying signal-to-noise ratios may result in non-stationary higher-order statistical properties that distort variance and AC1 and can lead to erroneous resilience estimates~\cite{smith2023reliability_ESD}. Additionally, time-correlated noise can introduce spurious changes in both variance and AC1 that are unrelated to actual changes in resilience~\cite{boers2021observation,boettner2022critical,morr2024detection}. To gauge the influence of such issues, Smith et al.~\cite{smith2023reliability} proposed jointly analyzing $\ELvar$ (a variance-based indicator) and $\ELac$ (an AC1-based indicator), arguing that their agreement provides a justification for their use. Deviations between these indicators were suggested as an uncertainty metric on the underlying modelling assumptions~\cite{ditlevsen2010tipping,smith2023reliability,chen2024accelerating}. Applying this framework to global vegetation dynamics, they found that the consistency between $\ELvar$ and $\ELac$ varies with biomass levels, with lower agreement observed in high-biomass regions such as tropical forests. However, a formal mathematical explanation for the relationship between $\ELvar$ and $\ELac$ remains absent, and the mechanisms underlying the link between their agreement and biomass levels are poorly understood.

In this study, we present a general analytical framework for understanding the relationship between the two most widely used resilience indicators $\ELvar$ and $\ELac$. We derive an analytical expression that characterizes their statistical dependence and reveals a fundamental sensitivity to the initial conditions of the time series. To investigate the impact of realistic data imperfections, we generate synthetic time series with controlled missing values and outliers, and assess their impact on the agreement between the resilience indicators $\ELvar$ and $\ELac$. We then apply this approach to satellite-derived vegetation indices across different land-cover types. Our results show that missing values tend to weaken indicator agreement, particularly in high-biomass systems, while outliers can systematically bias $\ELac$ downwards, leading to potential overestimation of resilience.

\section{Results} 
\subsection{Statistical dependence of resilience indicators}
We begin by reexamining the definition of the two most widely adopted resilience indicators, $\ELac$ and $\ELvar$, as:
\begin{eqnarray}\label{eq:lambda_ac1}
\ELac = & \log\left(\EAC\right),
\end{eqnarray}
and 
\begin{eqnarray}\label{eq:lambda_var}
    \ELvar = & \frac{1}{2}\log\left(1-\frac{\widehat{\sigma}_\varepsilon^2}{\EVar}\right),
\end{eqnarray}
respectively (see Methods). At first glance, Eq.~(\ref{eq:lambda_var}) suggests that $\ELvar$ primarily synthesizes information about the magnitude of fluctuations in both the noise term and the process $X$, i.e. $\widehat{\sigma}_\varepsilon^2$ and $\EVar$, respectively. However, $\widehat{\sigma}_\varepsilon^2$ in Eq.~(\ref{eq:lambda_var}), which is obtained by analyzing the residuals after performing a least squares regression, depends on AC1 according to the following formulation:
\begin{equation}\label{eq:dependent}
\widehat{\sigma}_\varepsilon^2=\frac{1}{N-1}\sum_{i=1}^{N-1}\left(X_{i+1}-\EAC\cdot X_i\right)^2.
\end{equation}
It is evident that $\widehat{\sigma}_\varepsilon^2$ depends on $\EAC$, suggesting that $\ELac$ and $\ELvar$ are statistically dependent. By inserting Eq.~(\ref{eq:dependent}) into Eq.~(\ref{eq:lambda_var}), we establish a direct functional relationship between these two resilience indicators (see Supplementary Materials for details), expressed as:
\begin{equation}\label{eq:lambda_relation}
    \ELvar=\frac{1}{2}\log\left(
    \underbrace{1-\frac{N}{N-1}\left(1- \exp\left(2\ELac\right)\right)}_{first \ term}+\underbrace{\frac{1}{N-1}\frac{X_1^2+\exp\left(2\ELac\right)X_N^2}{\EVar}}_{ second \ term}\right).
    \end{equation}
We note that in the absence of the second term inside the logarithm, Eq.~(\ref{eq:lambda_relation}) simplifies to a universal relationship that is independent of any specific realisation of the time series:
\begin{equation}\label{eq:lambda_relation-2}
\ELvar = \frac{1}{2}\log\left(
    1-\frac{N}{N-1}\left(1- \exp\left(2\ELac\right)\right)\right)
\end{equation}
This expression serves as a lower bound for the relationship between $\ELvar$ and $\ELac$, as the second term in Eq.~(\ref{eq:lambda_relation}) is non-negative. While the second term captures the influence of specific realisations of the time series, it depends only on the relative amplitude of the first and last data points (normalized by the average amplitude). Since $\ELac$ is negative, $\exp (2\ELac) X_N^2$ is generally negligible compared to $X_1^2$, rendering the second term primarily determined by $X_1^2/\EVar$ (see Supplementary Materials). Therefore, Eq.~(\ref{eq:lambda_relation}) confirms that the two resilience indicators, $\ELvar$ and $\ELac$, are not statistically independent. Instead, their relationship is largely determined by the relative amplitude of the first data point, regardless of the underlying system dynamics.

\begin{figure}[htp]
\noindent\includegraphics[width=1\textwidth]{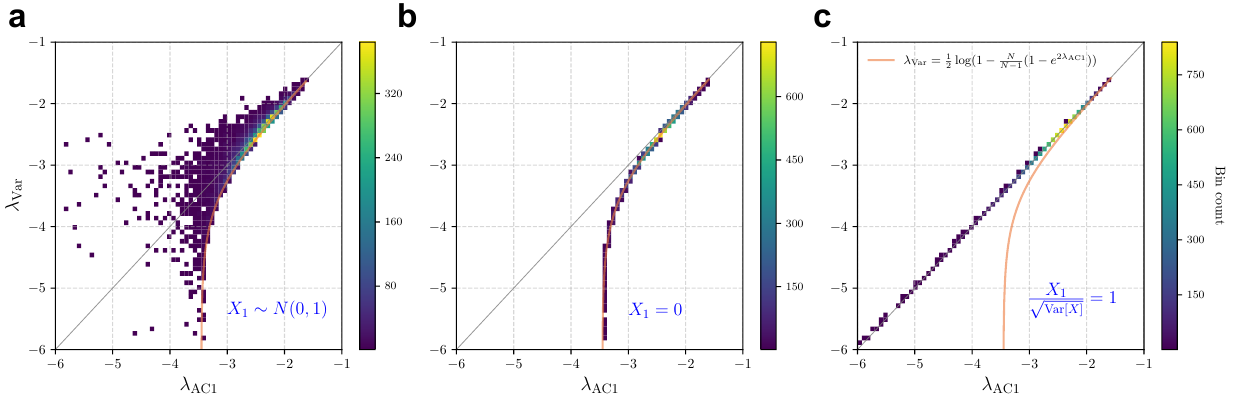}
\caption{\textbf{Relationship between resilience indicators $\ELvar$ and $\ELac$ under different initial conditions.} For each condition we generate $n = 10{,}000$ time series from an AR(1) process, each of length 1,000. Three scenarios are considered: (\textbf{a}) the first data point is unchanged, i.e $X_1 \sim N(0,1)$; (\textbf{b}) this first data point is artificially set as $X_1 = 0$; and (\textbf{c}) the first data point is set to satisfy $X_1/\sqrt{\EVar} = 1$. The orange curve represents the lower bound of the distribution, as characterized by the universal function given in Eq.~(\ref{eq:lambda_relation-2}).}
\label{fig:fig1}
\end{figure}

To numerically confirm the relationship established in Eq.~(\ref{eq:lambda_relation}), we generate a total of 10,000 time series from an AR(1) process with parameters $\alpha = e^{-2.5}$ and $\sigma_{\epsilon} = 1$ (see Methods). We then modify the first data point of each series to investigate its influence on the relationship between the reslience indicators $\ELvar$ and $\ELac$. Specifically, we consider three scenarios:

(1) The baseline scenario did not involve modifications, with the amplitude of the first data point following a normal distribution, i.e. $X_1 \sim N(0,1)$, representing expectations for a random initial measurement chosen from a stationary dynamic real-world system. As shown in Fig.~\ref{fig:fig1}a, the relationship between $\ELvar$ and $\ELac$ lies above the universal lower bound (the orange curve). This theoretical lower bound, described by Eq.~(\ref{eq:lambda_relation-2}), provides a quantitative constraint on the relationship between $\ELvar$ and $ \ELac$.

(2) The first data point is set to zero, i.e. $X_1 = 0$. Here, the second term on the right-hand side of Eq.~(\ref{eq:lambda_relation}) almost vanishes, causing the relationship between $\ELvar$ and $\ELac$ to follow the universal lower bound (Eq.~(\ref{eq:lambda_relation-2})). Consequently, as shown in Fig.~\ref{fig:fig1}b, data points from different time series collapse onto the orange curve.

(3) The relative amplitude of the first data point is set to unity, i.e. $X_1/\sqrt{\EVar} = \pm 1$, representing the statistically expected situation in a stationary process.
Under this constraint, Eq.~(\ref{eq:lambda_relation}) predicts an approximate equality between $\ELvar$ and $\ELac$ ($\ELvar \approx \ELac$), resulting in a clustering of data points along the identity line in the $\ELvar - \ELac$ scatter plot (Fig.~\ref{fig:fig1}c). 

These scenarios demonstrate that, in principle, a 1:1 relationship of $\ELvar$ and $\ELac$ can be expected, such as in the limit of an infinite number of data points $N$, or when taking the mean value across a huge number of realisations, since Eq.~(\ref{eq:lambda_relation}) reduces to $\ELvar = \ELac$ when $N \rightarrow \infty $. However, any point-wise realisation of $\ELac$ will inevitably exhibit scatter along the $\ELvar$ dimension, as illustrated in Fig.~\ref{fig:fig1}a. While this figure is based on time series generated by an AR(1) process, the underlying relationship is broadly applicable: Eq.~(\ref{eq:lambda_relation}) holds for any gapless time series, irrespective of the specific data-generating mechanism (see Supplementary Fig.~1 for additional examples). Notably, although $\ELvar$ and $\ELac$ are computed using the entire time series, their agreement is predominantly influenced by the first data point. Modifying only this initial value can significantly alter the relationship between the two indicators, even if all other properties of the time series remain unchanged. This sensitivity highlights the need for caution when interpreting strong correlations between these resilience indicators as sufficient evidence supporting the applicability of CSD analyses.

 \subsection{Missing values undermine the agreement of resilience indicators}

The statistical dependence between $\ELvar$ and $\ELac$, as expressed in Eq.~(\ref{eq:lambda_relation}), relies on the assumption that the time series contains no missing values (or data gaps). However, this assumption is frequently violated in real-world applications~\cite{bochow2025reconstructing}. In remote sensing datasets, for instance, missing values often result from the exclusion of spurious observations caused by factors such as cloud contamination in optical sensors, frozen ground conditions, or radio-frequency interference in radar systems~\cite{weiss2007lai,kandasamy2013comparison}. Missing values complicate the derivation of a general mathematical relationship between $\ELvar$ and $\ELac$, as the relationship becomes sensitive to the specific pattern and distribution of the gaps. To systematically assess the impact, we introduce artificial missing values into synthetic time series, allowing a controlled evaluation of their influence on CSD-based indicators.

We generate synthetic time series ($n=10{,}000$), each of length $N=1{,}000$, using an AR(1) process with parameters $\alpha = e^{-2}$ and $\sigma_{\epsilon} = 1$. Missing values are introduced by randomly removing $N_m$ data points from each series, where the missing value fraction is defined as $r = N_m/N$. Resilience indicators $\ELvar$ and $\ELac$ are computed using only available (non-missing) data points (see Supplementary Materials for the exact formulas). As illustrated in Fig.~\ref{fig:fig2}, the presence of missing values substantially alters the relationship between $\ELac$ and $\ELvar$. Even a small missing value fraction ($ r = 1\%$; Fig. \ref{fig:fig2}a-c) can significantly distort the scatter plots compared to the gap-free case (Fig.~\ref{fig:fig1}). As the fraction of missing data increases, the distinctions among the three initial conditions diminish, and the universal function (orange curve) no longer acts as a lower bound. These observations demonstrate that Eq.~(\ref{eq:lambda_relation}) does not hold in the presence of missing values. Notably, increasing missing value fraction from $r = 1\%$ to $r =20\%$ results in greater dispersion in the scatter plots and a marked decrease in the correlation between the two indicators. This reduced agreement arises because missing values affect the estimation of AC1 and variance in distinct ways. Specifically, a single missing value removes two consecutive data points from the AC1 calculation, whereas only one data point is excluded from the variance estimate. As a result, AC1 and variance are computed from different subsets of the data, introducing inconsistencies that propagate into $\ELac$ and $\ELvar$. These differential effects are further illustrated in Supplementary Fig.~2.

\subsection{Real-world impact on high-biomass regions}

To illustrate how missing values affect resilience estimation in real-world settings, we use global vegetation datasets derived from the Moderate Resolution Imaging Spectroradiometer (MODIS) as a case study. In particular, we focus on five vegetation indices: NDVI, kNDVI, EVI, GPP, and LAI (see Methods). This analysis also helps to explain a previously reported phenomenon by Smith et al.~\cite{smith2023reliability}, who observed that the relationship between $\ELvar$ and $\ELac$ varies substantially across land-cover types. We propose that this variation is substantially affected by differences in the fraction of missing values across ecosystems.

On a global scale, the MODIS data reveal a strong association between the fraction of missing values and regional biomass. As illustrated by the purple lines in Fig.~\ref{fig:fig3}, data from high-biomass land-cover types, such as evergreen forests, consistently exhibit a greater share of missing observations. In contrast, data from low-biomass land-cover types, such as open shrublands, experience lower fractions of missing values. This spatial correspondence is further supported by the similarity between global biomass distributions (Supplementary Fig.~3) and the missing values distribution of the MODIS NDVI dataset (Supplementary Fig.~4). One key cause of this pattern lies in atmospheric conditions: high static stability (especially in the sinking branch of the Hadley cell in the subtropics) is associated with low cloud coverage, low precipitation, and hence low biomass, whereas tropical, high-biomass regions typically exhibit high cloud coverage   ~\cite{spracklen2012observations,duveiller2021revealing}. Moist and cloudy atmospheric conditions obstruct optical and thermal satellite sensors, leading to more data loss compared to arid regions. 

Given that missing values weaken the agreement between $\ELvar$ and $\ELac$, their prevalence in high-biomass ecosystems likely contributes to the reduced indicator agreement observed in these regions~\cite{smith2023reliability}. This interpretation is supported by a strong negative correlation between the missing value fraction (solid purple lines in Fig.~\ref{fig:fig3}) and the agreement between $\ELvar$ and $\ELac$ (solid green lines in Fig.~\ref{fig:fig3}) across various land-cover types. This pattern is robust across all five MODIS indices considered, underscoring a strong link between missing value fraction and reduced agreement between resilience indicators.

We further assess the influence of missing values on the agreement between $\ELvar$ and $\ELac$ in remote sensing data through a series of synthetic experiments. For each MODIS vegetation dataset, we generate a corresponding synthetic dataset using an AR(1) model with parameters $\alpha = e^{-1}$ and $\sigma_{\epsilon} = 1$. These synthetic datasets match the remote sensing datasets in both the number and length of time series (n = 10,000 for each natural land-cover type). Critically, we ensure that the proportion of missing values for each land-cover type in the synthetic datasets mirrors that observed in the remote sensing data. For example, the evergreen vegetation class in the synthetic NDVI dataset maintains the same proportion of missing values as observed in the MODIS NDVI dataset for evergreen vegetation. This design guarantees that any difference observed in resilience indicators arises solely from missing value patterns, rather than from other data characteristics or underlying dynamic processes. As shown in the Fig.~\ref{fig:fig3}b,d,f,h,j, the correlation between $\ELvar$ and $\ELac$ in the synthetic datasets (dotted blue lines) exhibits strong agreement with the results computed from the corresponding remote sensing datasets (solid green line). This agreement holds true for different MODIS vegetation indices with different spatial resolutions (Supplementary Fig.~5) which confirms that the observed divergence of the
$\ELvar$ - $\ELac$ relationship across land-cover types is primarily driven by differences in the fraction of missing values. We emphasize that we can broadly reproduce this divergence using only the missing value fraction, and ignoring underlying differences in different ecosystems (e.g., forest vs savannah). This implies both a very strong control on resilience estimates by missing values, and a strong underlying similarity in the dynamics of different vegetated ecosystems. To ensure this result is not an artifact of the deseasoning method, we repeated the analysis  using alternative methods (e.g. STL decomposition), which yield consistent results (Supplementary Fig.~6), reinforcing the robustness of this conclusion.

 \subsection{Outliers introduce systematic biases in resilience assessments}
Beyond missing values, our theoretical analysis in the form of Eq.~(\ref{eq:lambda_relation}) and Fig. \ref{fig:fig1} also suggests that outliers may introduce additional systematic biases in resilience indicators, because they can affect the variance ratio between the first data point and the rest of the data. Such biases are particularly relevant for satellite-derived products, which are often affected by spurious data points that deviate from expected annual or seasonal patterns~\cite{zhang2007anomaly}. These outliers may arise from various sources, including atmospheric effects (e.g., cloud cover, aerosol interference causing over- or underestimation of vegetation greenness)~\cite{frey2008cloud}; sun angle and topographic shadow effects (e.g., shadowing leading to artificially low NDVI values in mountainous terrain)~\cite{matsushita2007sensitivity}, and instrument limitations (e.g., receiver sensitivity to connectivity, weather, topography, and canopy interference)~\cite{knott2023effects}, among others. Furthermore, as CSD-based analyses typically rely on anomaly time series obtained after detrending and deseasonalizing the data, inappropriate or suboptimal preprocessing methods can also introduce spurious outliers~\cite{smith2023reliability}.

For time series without missing values, the relationship between $\ELvar$ and $\ELac$, as described by Eq.~(\ref{eq:lambda_relation}), is governed by the relative amplitude of the first data point, $X_1^2/\EVar$. Consequently, the influence of outliers can be understood under two conditions. First, an outlier occurring at the first data point ($X_1$) directly modifies the relationship between the two indicators. Second, when outliers are present elsewhere in the time series, particularly if they are numerous or of large magnitude, they increase the overall variance, thereby diminishing the relative weight of $X_1^2/\EVar$. In this case, the relationship between $\ELvar$ and $\ELac$ tends to converge toward the universal lower bound defined by Eq.~(\ref{eq:lambda_relation-2}). This behaviour is confirmed numerically in Supplementary Fig. 7.

For time series with missing values, however, the influence of outliers on $\ELvar$ and $\ELac$ cannot be understood in such an algebraic way. Missing values mean that there are, in effect, several time series points at the beginning and end of the interrupted time series parts. These all play into the relation between $\ELvar$ and $\ELac$ in an algebraically intransparent way. We can \textit{a priori} not expect any specific change in correlation between the two quantities with respect to an outlier in the first data point. Nonetheless, we can investigate these effects through synthetic time series experiments.

We generate time series using an AR(1) process with parameters $\alpha = e^{-2}$ and $\sigma_{\epsilon} = 1$. Five percent of the values in each series are randomly designated as missing, while additional five percent are assigned as artificial outliers. These outliers are generated by randomly sampling from the intervals $[ \mu + k\sigma, \mu + (k+1)\sigma ]$ or $[ \mu - (k+1)\sigma, \mu - k\sigma ]$, where $\mu$ and $\sigma$ represent the mean and standard deviation of each time series, respectively. The parameter $k$ controls the magnitude of deviation, with larger values producing more extreme deviations from the mean. An example of artificial outliers is illustrated in Supplementary Fig.~8. The magnitude of outliers is quantified by computing the kurtosis of the dataset (see Supplementary Materials), a widely adopted statistical measure for outlier detection~\cite{livesey2007kurtosis, westfall2014kurtosis}. By varying $k$, we assess the impact of increasing outlier magnitude on the performance of resilience indicators. As shown in Fig.~\ref{fig:fig4}, higher outlier magnitudes, as reflected in increased kurtosis, lead to a reduced correlation between the two resilience indicators. Notably, the presence of outliers substantially alters the distributional properties of $\ELvar$ and $\ELac$. In the baseline case, where time series are affected only by missing values and contain no outliers (Fig.~\ref{fig:fig4}a), the scatter plot of $\ELvar$ versus $\ELac$ is evenly distributed around the 1:1 line, forming what we refer to as the ``balanced pattern". In this pattern, the mean position of the scatter points (red cross) lies on the 1:1 line, indicating that the average deviation between $\ELvar$ and $\ELac$ across the dataset is close to zero. This suggests that, despite point-wise variability, averaging resilience indicators over the full dataset still yields a consistent estimate of system resilience. As outlier magnitude increases (Fig.~\ref{fig:fig4}b–f), the distribution progressively shifts toward the upper left of the 1:1 line, forming a “biased pattern.” Here, the mean position of the scatter points moves above the 1:1 line, indicating a systematic divergence in which $\ELac$ is consistently lower than $\ELvar$. As a result, averaging the two resilience indicators across the dataset (e.g., the red cross in Fig.~\ref{fig:fig4}f) yields an inconsistent estimate of system resilience, reflecting the distortion introduced by outliers.

To investigate the mechanism behind this divergence, we separately examine the sensitivity of each indicator to outlier magnitude. We find that $\ELvar$ remains relatively stable to increasing outlier magnitude. This stability arises from its formulation in Eq.~(\ref{eq:lambda_var}): although both $\widehat{\sigma}_\varepsilon^2$ and $\EVar$ increase with stronger outliers, their ratio remains relatively stable (see Supplementary Fig.~9). In contrast, $\ELac$ decreases as outlier magnitude increases. This decline reflects the disruption of autocorrelation caused by randomly placed outliers, which weaken the correlation between successive values and thus reduce AC1-based resilience estimates. As a result, outliers systematically impose a negative bias on $\ELac$ (a shift to the left in Fig.~\ref{fig:fig4}). This leads to an overestimation of resilience when relying on the AC1 indicator and, thus, potentially to late warning of forthcoming transitions.

This transition from balanced to biased patterns is also observable in satellite-derived vegetation indices. Specifically, time series from evergreen broadleaf forests display a biased pattern, whereas those from closed shrublands exhibit a balanced pattern (see Supplementary Fig.~10 or Fig. 2a,b in Ref.~\cite{smith2023reliability}). These differences align with the significantly higher kurtosis values observed in evergreen forests compared to shrublands (Fig.~\ref{fig:fig5}a), suggesting that variations in outliers contribute to the shift in the distributional properties of $\ELvar$ and $\ELac$ between different land-cover types.

To further demonstrate that the distributional shift in resilience indicators between evergreen broadleaf forests and closed shrublands arises from data imperfections, we reproduced this phenomenon using synthetic time series by incorporating the observed differences in missing values and outliers between these two land-cover types. As shown in Fig.~\ref{fig:fig5}b,c, synthetic time series generated from an AR(1) process with parameters $\alpha = e^{-1}$ and $\sigma_{\epsilon} = 1$ exhibit distinct indicator patterns depending on the level of data imperfection. Specifically, a balanced pattern (Fig.~\ref{fig:fig5}b) emerges under low fractions of missing values and outlier magnitude, matching the fraction of missing values = 1.6\% and median kurtosis = 3.81 observed in MODIS NDVI dataset for closed shrublands. In contrast, a biased pattern (Fig.~\ref{fig:fig5}c) arises under more severe data imperfections, with the fraction of missing values = 66.0\% and median kurtosis = 7.17 aligned with those found in evergreen broadleaf forests. The close agreement between the synthetic (Fig.~\ref{fig:fig5}b,c) and empirical results (Supplementary Fig.~10) confirms that heterogeneity in data quality, particularly differences in missing value proportions and outlier magnitude, plays a central role in shaping the relationship between resilience indicators across land-cover types in remote sensing observations.

\section{Discussion and Conclusion}

Persistent warming and anthropogenic pressures are reducing the resilience of Earth system components and especially the Earth's ecosystems, raising the risk of abrupt regional shifts~\cite{lenton2024remotely}. Detecting early warning signals of instability is therefore urgent~\cite{forzieri2022emerging}. CSD-based indicators offer an alternative to disturbance-based measurements, enabling global resilience assessments and tracking stability changes across diverse domains~\cite{verbesselt2016remotely,smith2022empirical,boers2021observation,boers2021critical}. To mitigate spurious signals from individual indicators, recent studies have integrated multiple resilience indicators into composite frameworks, yielding more reliable assessments~\cite{dakos2022ecological,smith2023reliability,smith2023reliability_ESD}. However, the statistical dependencies between early-warning indicators -- and their sensitivity to common data issues -- remain poorly understood. To address this, we derive a general mathematical framework linking two widely used CSD-based resilience indicators, $\ELac$ and $\ELvar$, which allows us to relate their agreement to data quality. Using synthetic and satellite-derived vegetation data, we show that missing values and outliers systematically undermine indicator consistency, offering new guidance for robust resilience assessments.

Our mathematical analysis reveals that, for complete time series (i.e. without missing values), the relationship between $\ELac$ and $\ELvar$ is fully determined by a functional dependence governed primarily by the amplitude of the first data point relative to the overall variance (Eq.~(\ref{eq:lambda_relation})). This finding underscores a previously unacknowledged sensitivity: the apparent agreement or disagreement between the two resilience indicators may arise solely from properties of the first data point in a time series rather than reflecting underlying CSD dynamics. As such, high consistency between $\ELac$ and $\ELvar$ should be interpreted with caution, as it does not inherently guarantee the appropriateness of the data for CSD analysis. As mentioned above, any dataset, even if not generated by an autoregressive process, can show such consistency between $\ELac$ and $\ELvar$. These findings challenge the assumption that such indicators provide independent confirmation of resilience changes~\cite{smith2023reliability,chen2024accelerating,wu2024alteration,liu2025accelerated} and underscore the importance of accounting for initial conditions when interpreting the coherence between indicators. Moreover, the potential interdependence between other resilience indicators, such as those based on estimating the drift function of a stochastic model~\cite{morr2024anticipating}, warrants further investigation.

\begin{figure}[htp]
\noindent\includegraphics[width=1\textwidth]{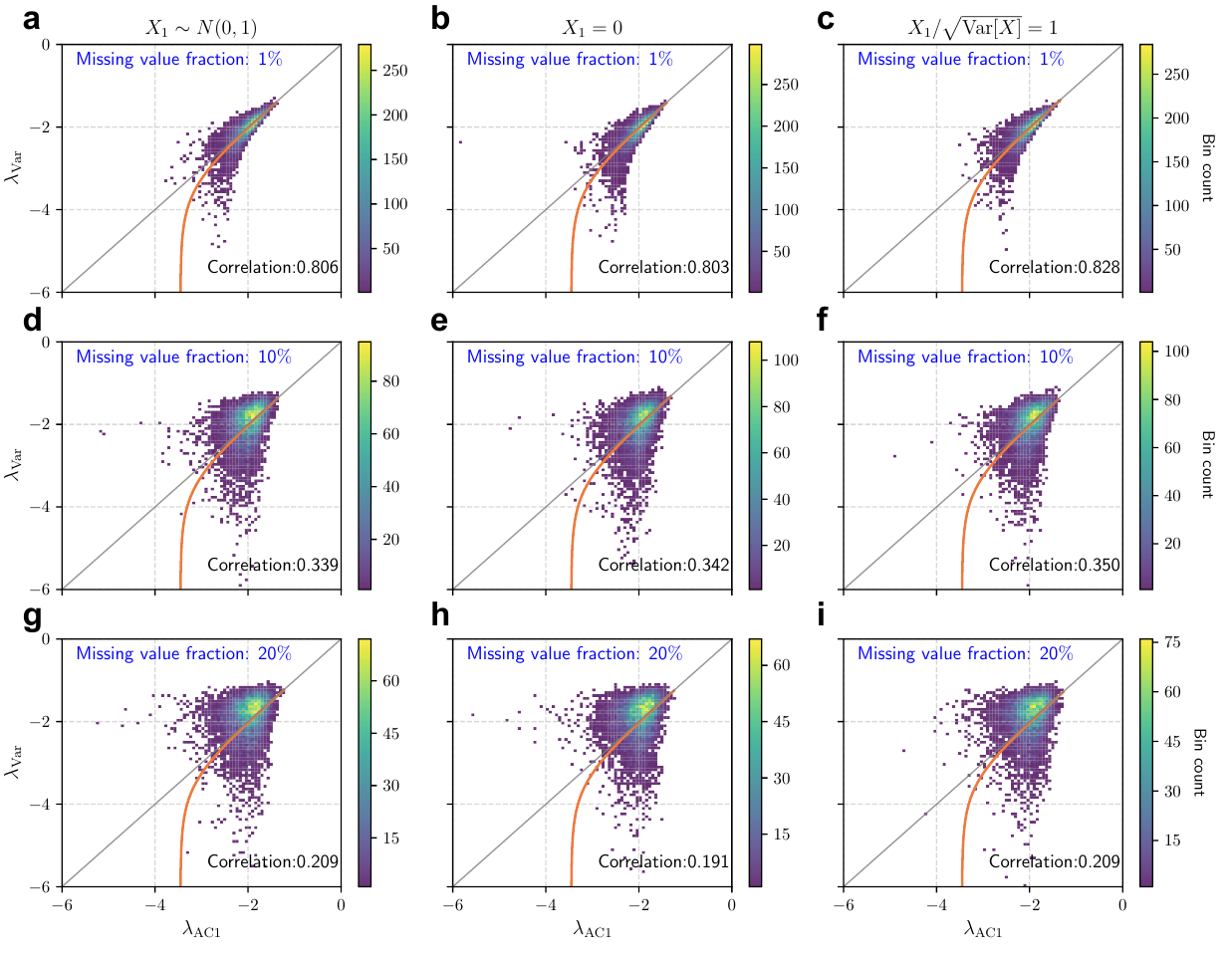}
\caption{\textbf{Relationship between $\ELvar$ and $\ELac$ under varying fractions of missing values.} Each condition consists of $n$=10,000 time series generated from an AR(1) process, each of length 1,000. Missing data were introduced randomly with fractions of 1\% (\textbf{a-c}), 10\% (\textbf{d-f}), and 20\% (\textbf{g-i}). The columns correspond to three distinct initial value ($X_1$) settings, consistent with the treatments presented in Fig.~\ref{fig:fig1}. The orange curve represents the universal function described by Eq.~(\ref{eq:lambda_relation-2}), which no longer serves as the lower bound when the time series contains missing values.}
\label{fig:fig2}
\end{figure}

\begin{figure}[htp]
\noindent\includegraphics[width=1\textwidth]{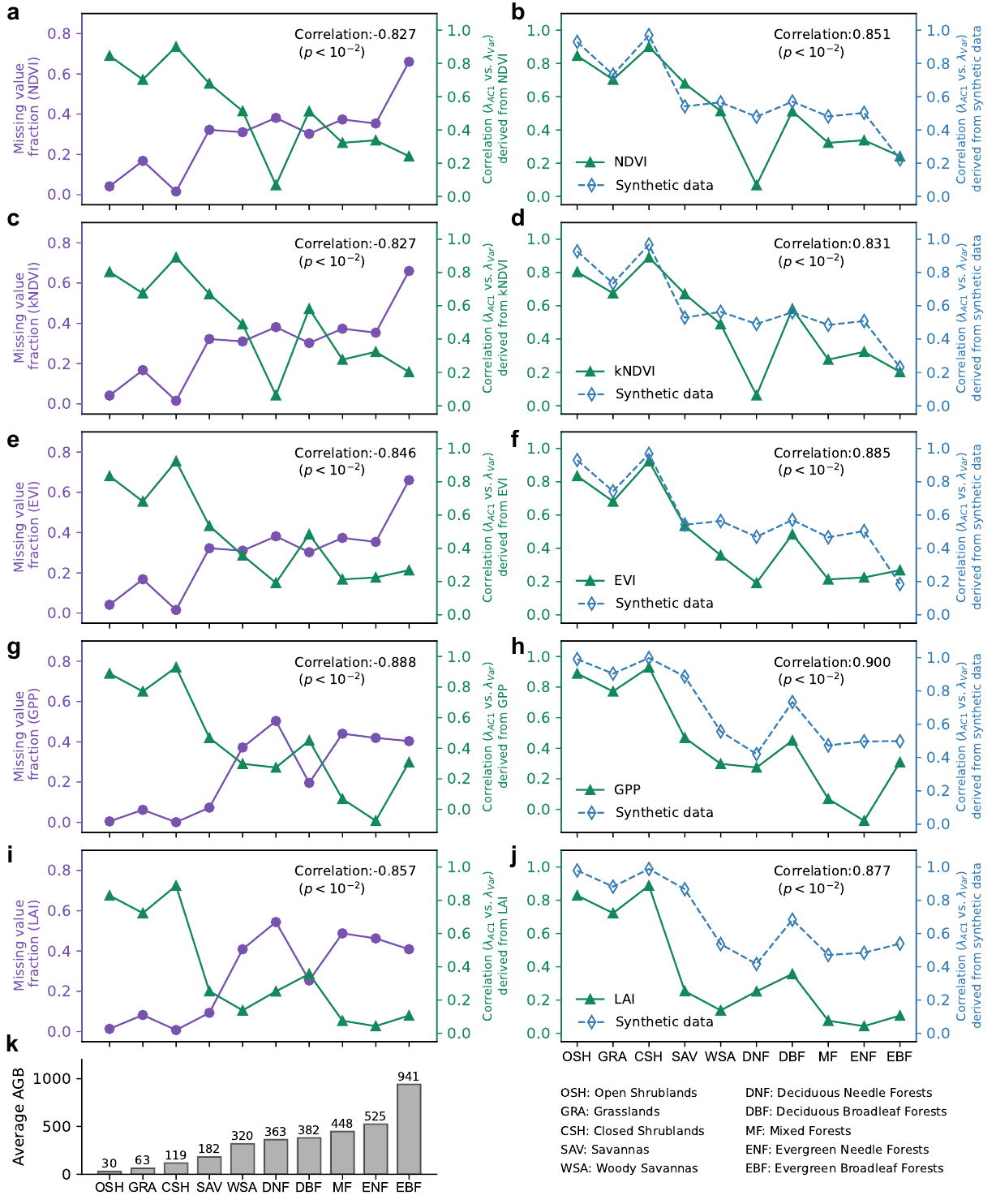}
\caption{\textbf{Relationship between resilience indicator consistency (Correlation between $\ELvar$ and $\ELac$) and missing value ratios across land-cover types in MODIS and synthetic datasets.} \textbf{a,c,e,g,i,} Resilience indicator consistency is calculated from $n = 10{,}000$ points per natural land-cover type using MODIS vegetation indices at native sensor resolutions (EVI, NDVI, kNDVI: 250 m; GPP, LAI: 500 m) (solid green lines). Median fraction of missing values across land-cover types for all MODIS vegetation indices are represented by the solid purple lines. \textbf{b,d,f,h,j,} Dotted blue lines represent resilience indicator consistency derived from synthetic datasets, generated using an AR(1) process that incorporates the fraction of missing values observed in the corresponding MODIS land-cover types. Pearson correlation coefficients between the solid purple and green lines, and between solid green and dotted blue lines, are indicated in each panel. \textbf{k,} Histogram of average above-ground biomass (AGB) density across ten land-cover types. All panels share the same  \textit{x}-axis labels, representing land-cover types ordered by decreasing average AGB.} 
\label{fig:fig3}
\end{figure}

We further assess the impact of missing values on resilience indicators. Introducing artificial missing values into synthetic time series reveals a strong decline in indicator agreement as the fraction of missing values increases (Fig.~\ref{fig:fig2}). Applying this framework to MODIS vegetation data, we find a robust negative correlation between missing value frequency and indicator agreement across land-cover types (Fig.~\ref{fig:fig3}). This relationship offers a clear explanation for the divergence in indicator agreement between high- and low-biomass areas reported by Smith et al.~\cite{smith2023reliability}. Rather than reflecting inherent ecological differences, this discrepancy appears to be primarily due to data quality issues. High-biomass ecosystems tend to experience more frequent cloud cover, resulting in higher rates of missing values in remote sensing data. Our findings demonstrate that these missing values could introduce biases in mean vegetation resilience estimates.

\begin{figure}[htp]
\noindent\includegraphics[width=1\textwidth]{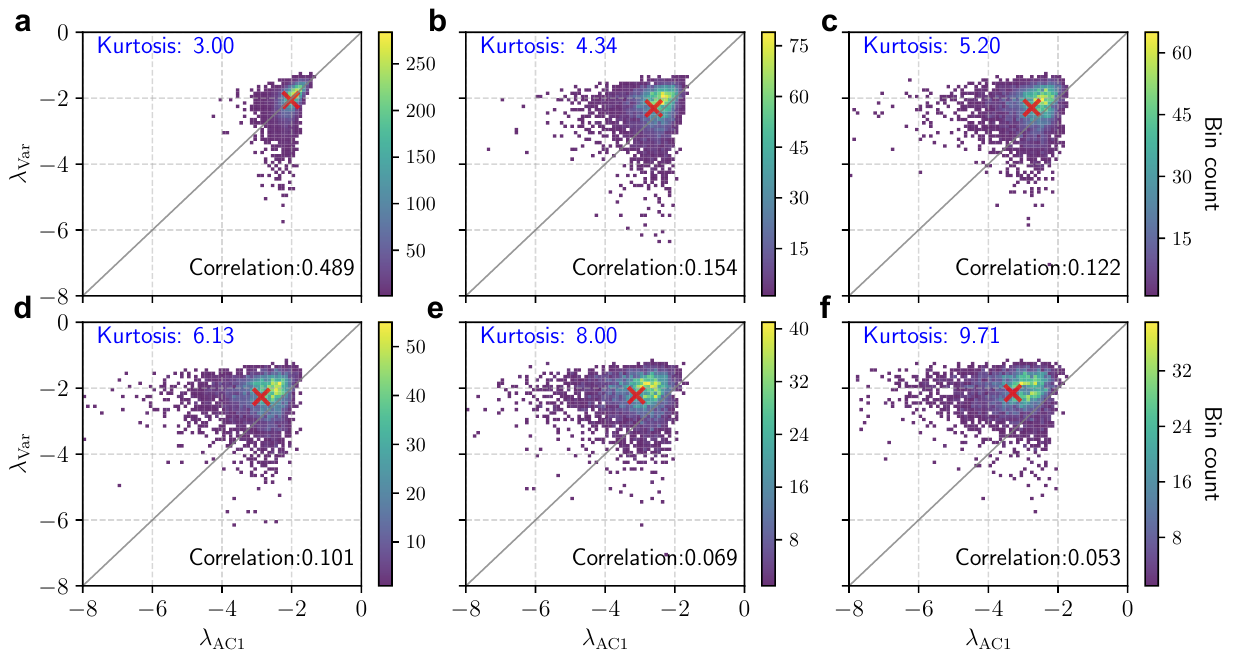}
\caption{\textbf{Relationship between $\ELvar$ and $\ELac$ under varying outlier magnitudes.} Each panel shows results from $n = 10{,}000$ simulated time series generated from an AR(1) process, each of length $1{,}000$. Five percent of the values are randomly designated as missing, and an additional five percent are set as outliers. \textbf{a,} Baseline case with no outliers. \textbf{b-f,}
Outlier magnitude increases progressively by scaling deviations with parameter $k$ ($k =3.0,3.5,4.0,5.0,6.0$), resulting in increasing kurtosis across panels. Red crosses indicate the mean position of the scatter plot in each panel.}
\label{fig:fig4}
\end{figure}

\begin{figure}[htp]
\noindent\includegraphics[width=1\textwidth]{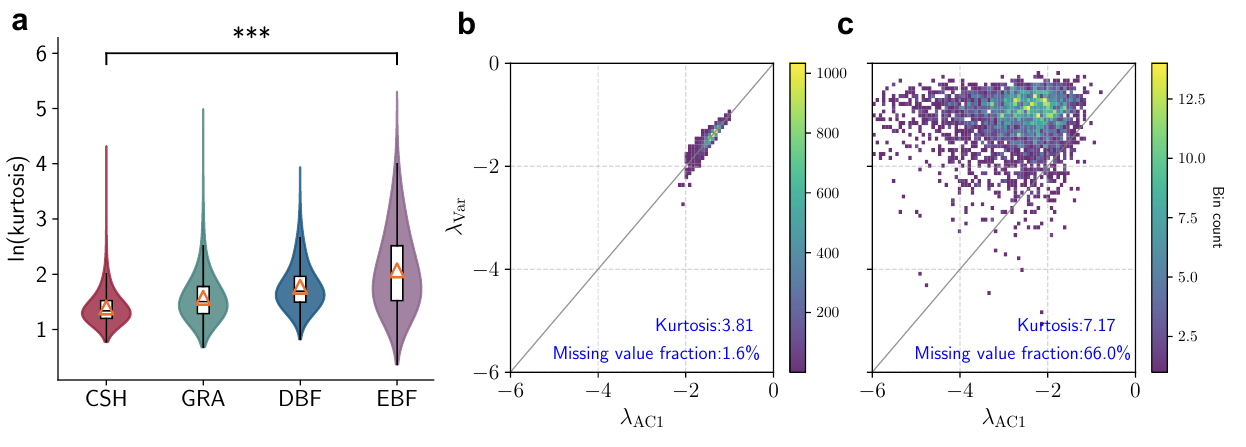}
\caption{\textbf{Outlier and missing value induced shifts in the relationship between resilience indicators.} \textbf{a} Kurtosis of (de-seasonalized and de-trended) NDVI data. The central line denotes the median, the lower and upper hinges represent the 25th and 75th percentiles, respectively, whiskers extend to the 95\% confidence intervals, and triangles indicate the mean. Significant differences in kurtosis ($p<0.001$) between closed shrublands and evergreen broadleaf forests, identified by ANOVA, are marked with asterisks. \textbf{b, c} Influence of missing values and outliers on the distributional properties of $\ELvar$ and $\ELac$. Simulated time series demonstrate a shift from a balanced pattern (\textbf{b}) to a biased pattern (\textbf{c}) as both the fraction of missing values and the magnitude of outliers increase. Outliers were introduced by designating 4\% of data points as extreme values, with the outlier magnitude parameter $k$ increasing from 2.67 (\textbf{b}) to 8.50 (\textbf{c}).
The fraction of missing values and kurtosis values in \textbf{b} and \textbf{c} match those observed in MODIS NDVI data for shrublands (fraction of missing values = 1.6\%, median kurtosis = 3.81) and evergreen broadleaf forests (fraction of missing values = 66.0\%, median kurtosis = 7.17), respectively.}
\label{fig:fig5}
\end{figure}

Alongside missing values, outliers represent another significant data quality issue affecting resilience indicators. Beyond reducing consistency, outliers can alter the distributional properties of $\ELac$ and $\ELvar$, as evidenced by changes in the shape of their scatter plots (Fig.~\ref{fig:fig4} and \ref{fig:fig5}). Notably, high-magnitude outliers systematically bias AC1-based indicators toward overestimating resilience, with these distortions particularly evident in vegetation observations. Comparing resilience indicators derived from remote sensing data reveals significant performance differences between evergreen forests (characterized by high outliers and frequent missing values) and open shrublands (where both issues are less prevalent). This suggests that AC1-based metrics may systematically overestimate vegetation resilience in evergreen-dominated ecosystems, such as tropical forests. 

While our analysis focuses on the fraction of missing values and outlier magnitudes, other data characteristics may also influence resilience estimates and warrant futher investigation. For instance, the distribution pattern of missing values (see Supplementary Fig.~11) can introduce additional biases~\cite{ben2023uncertainties}, increasing the number of outliers (in contrast to their magnitude, as shown above) can further reduce the agreement between resilience indicators (see Supplementary Fig.~12), and different mechanisms for generating outliers in real systems could lead to varying impacts. Although the relationship between $\ELac$ and $\ELvar$ provides valuable insights into data quality, it should not be regarded as a definitive measure of a dataset’s suitability for CSD analysis. Instead, we propose that comparing CSD-based resilience indicators with empirically derived recovery rates following large perturbations offers a more robust framework to evaluate the suitability of CSD methods~\cite{smith2023reliability,smith2022empirical}. Nevertheless, both $\ELac$ and $\ELvar$ remain powerful tools for quantifying resilience in dynamical systems, provided their limitations are carefully considered. 

Our results underscore the importance of continuous, high-quality datasets for reliable resilience assessment. Although remote sensing and other observational platforms hold great promise for global-scale monitoring, they are often limited by noise, missing data, and irregular sampling -- issues that are especially pronounced in high-biomass and biodiverse regions that are vital to planetary resilience. Addressing data gaps while preserving key dynamical properties, such as AC1, remains a major challenge. Traditional gap-filling methods, including temporal resampling and linear interpolation, can introduce systematic biases in the higher-order statistics underlying resilience indicators even if the mean is unbiased \cite{smith2023reliability_ESD,ben2023uncertainties,smith2023reliability}. For example, gap filling by temporal resampling is sensitive to the relationship between the resampling window and the intrinsic timescale of the dynamics, potentially blending multiple lag structures. Similarly problematic, linear interpolation may impose artificially smooth dynamics, obscuring true variability. Emerging AI-based reconstruction methods offer a promising alternative~\cite{bochow2025reconstructing}, but their ability to recover the underlying system dynamics -- essential for resilience assessments -- remains uncertain and deserves further exploration.

In conclusion, this work bridges a critical gap between theoretical resilience frameworks and their empirical application, revealing the intricate interplay between data characteristics and statistical estimators. By formalizing the relationship between $\ELac$ and $\ELvar$ and diagnosing the impacts of missing values and outliers, we show how data quality can compromise the robustness of resilience assessments. Although vegetated ecosystems serve as a primary application in our study, the implications of our findings extend broadly to any context where resilience or stability is inferred from time series data.

\clearpage

\clearpage

\clearpage

\clearpage

\clearpage

\clearpage

\section{Methods}
\subsection{Time-series estimators for resilience}

We here follow Ref.~\cite{smith2023reliability} in deriving autocorrelation- and variance- based resilience indicators. In the framework of local resilience, i.e. resilience against small disturbances, the fundamental quantity of interest is the linear restoring rate, here termed $\lambda$. This is because the dynamics of any system close to a point attractor can be reduced to a linear system. If the internal disturbances of the system are modelled as Gaussian white noise, we obtain the Ornstein–Uhlenbeck process,
\begin{equation}
    \dint X_t=\lambda X_t\dint t+\sigma\dint W_t,
\end{equation}
where the variable $X_t$ represents deviations from the equilibrium of, e.g., a vegetation property. The linear restoring rate is negative ($\lambda \textless 0$) for stable dynamics, and increasing (decreasing) recovery rate $\lambda$ indicates a loss (gain) of resilience. The Gaussian white noise term $\dint W_t$ is represented by the differential of the Wiener process $W_t$.

In order to apply the theory to observational time series data, we discretized the above Ornstein–Uhlenbeck process into equal time steps of size $\Delta t$, which yields the characteristic order-one auto-regressive process (AR(1) process)
\begin{equation}\label{eq:ar1_process}
    X_{t+1}=\alpha X_t+\sigma_\varepsilon\varepsilon_t,
\end{equation}
where $\alpha=\exp(\lambda \Delta t)$ and $\varepsilon_t$ are i.i.d. unit normal random variables. The noise variance $\sigma_\varepsilon^2$ is given by:
\begin{equation}\label{eq:ar1_noise}
\sigma_\varepsilon^2 = -\frac{\sigma^2}{2\lambda}(1 - \exp(2\lambda \Delta t))
\end{equation}
Without loss of generality, we assume $\Delta t=1$ in the following for simplicity.

\subsubsection{Resilience estimator based on AC1}
A commonly used approach for estimating $\lambda$ in Eq.~(\ref{eq:ar1_process}) is based on the lag-1 autocorrelation~\cite{held2004detection,dakos2008slowing}. For the above AR(1) process, the theoretical lag-1 autocorrelation (AC1) is given by $\alpha=\exp(\lambda)$. With this relationship, we can estimate $\lambda$ as
\begin{eqnarray}
\ELac = & \log\left(\EAC\right),
\end{eqnarray}
where $\EAC$ represents the empirical estimator of the lag-1 autocorrelation computed from the observed time series~\cite{smith2023reliability}. 

\subsubsection{Resilience estimator based on variance}
An alternative approach to estimating $\lambda$ relies on the variance of the above AR(1) process~\cite{smith2023reliability}, which is given by: 
\begin{equation}\label{eq:var}
    \mathrm{Var}[X]=\frac{\sigma^2}{2\lambda}.
\end{equation}
By combining Eq.~(\ref{eq:var}) and Eq.~(\ref{eq:ar1_noise}) and eliminating $\sigma^2$, we obtain the following estimator for $\lambda$:
\begin{eqnarray}
    \ELvar = & \frac{1}{2}\log\left(1-\frac{\widehat{\sigma}_\varepsilon^2}{\EVar}\right).
\end{eqnarray}
Here, $\widehat{\sigma}_\varepsilon^2$ is estimated from a linear regression of the AR(1) process. Notably, the two estimators for the linear restoring rate, $\ELac$ and $\ELvar$, are not independent, a topic explored in detail in Section 3.1.

\subsection{Satellite data}
\subsubsection{Vegetation indices}
To assess the consistency of resilience indicators across global vegetation ecosystem, we utilize five vegetation indices derived from MODIS products in this study: (1) Enhanced Vegetation Index (EVI) and (2) Normalized Difference Vegetation Index (NDVI) from MOD13Q1 (250 m spatial resolution, 16-day composites)~\cite{didan2021modis}, (3) Gross Primary Productivity (GPP; MOD17A2, 500 m spatial resolution, 8-day composites)~\cite{running2015mod17a2h}, (4) Leaf Area Index (LAI; MCD15A3H, 500 m spatial resolution, 4-day composites)~\cite{myneni2021modis}, and (5) kernel-normalized difference vegetation index (kNDVI), a nonlinear extension of NDVI, which is calculated as
\begin{eqnarray}
    \mathrm{kNDVI} = \tanh{(\mathrm{NDVI}^2)}
\end{eqnarray}
Compared to conventional NDVI indices, kNDVI provides superior handling of nonlinearity, enhanced noise resistance, and improved temporal-spatial stability~\cite{camps2021unified}. All vegetation datasets are accessible via Google Earth Engine~\cite{gorelick2017google}, and only data points flagged as `highest quality' are used in our analysis. To maintain alignment with Ref.~\cite{smith2023reliability}, we set the study period to 2000-2022 (2002–2022 for LAI).

\subsubsection{Land cover data}

We use MODIS land-cover data~\cite{friedl2015mcd12c1} (MCD12Q1, 500 m spatial resolution, 2001–2021, annual) to subdivide results by land-cover type and mask out non-vegetated areas (e.g., water bodies, urban regions), based on Land Cover Type 1. Additionally, to minimize the influence of anthropogenic activity and ecosystem transitions on our results, we exclude any pixels that experienced a land-cover change (e.g., from forest to agriculture or agriculture to forest) at any point during 2001–2021. The land-cover data were resampled to a 250 m resolution to match the vegetation datasets using nearest-neighbor resampling.

We ensure equal representation of all land-cover types when comparing the consistency of resilience indicators by using a stratified random sample of 100,000 locations, evenly distributed across the ten relevant natural land-cover types based on the classification of International Geosphere-Biosphere Programme type 1~\cite{sulla2019hierarchical}. The spatial distribution of one realization of the stratified random sample is shown in Supplementary Fig.~3. We also used a global above-ground biomass density estimate from the 2010 composite~\cite{spawn2020harmonized} to assess the relationship between the consistency of resilience indicators and biomass across land cover types.

\subsubsection{De-Trending and De-Seasoning}
All CSD-based resilience indicators rely on perturbations of the state variable around its equilibrium. Therefore, the analyzed time series must be approximately stationary, requiring careful removal of long-term trends and seasonal signals~\cite{bathiany2024ecosystem}. To achieve this, we apply a rolling mean detrending method, followed by the removal of a third-order harmonic function fitted to the data for deseasoning — a technique shown to be particularly effective for processing remote sensing vegetation data~\cite{smith2023reliability}. As an alternative detrending and deseasoning approach, we also apply the widely used Seasonal-Trend decomposition using LOESS (STL)~\cite{cleveland1990stl} to cross-check our results.

\clearpage
\section*{Declarations}
\backmatter

\bmhead{Author contribution}
T.L., A.M., S.B., T.S. and N.B. conceived and designed the study. T.L. performed the simulation and computations, and analyzed the results. All authors discussed the results. T.L. wrote the paper with contributions from all authors.

\bmhead{Competing interests}

The authors declare no competing interests.

\bmhead{Supplementary information}

Supplementary information is available for this paper.

\bmhead{Data availability}
The MODIS datasets used in this study are all available at \url{https://modis.gsfc.nasa.gov/}, and can be accessed offline or via Google Earth Engine. The above-ground biomass density estimate~\cite{spawn2020harmonized} used in this study is available at \url{https://daac.ornl.gov/cgi-bin/dsviewer.pl?ds_id=1763}. Synthetic data can be reproduced via codes that will be publicly available upon acceptance of this manuscript for publication.

\bmhead{Code availability}
Python scripts used for deseasoning, detrending, and exporting MODIS vegetation data, as well as the code for reproducing the synthetic data used in this study, will be publicly available upon acceptance of this manuscript for publication.

\bmhead{Acknowledgements}
N.B. and S.B. acknowledge funding from the Volkswagen Foundation.
This is ClimTip contribution \#X; the ClimTip project has received funding from the European Union’s Horizon Europe research and innovation programme under grant agreement no. 101137601. 
This study received support from the European Space Agency Climate Change Initiative (ESA-CCI) Tipping Elements SIRENE project (contract no. 4000146954/24/I-LR). T.S. acknowledges support from the DFG STRIVE project (SM 710/2-1).


\end{document}